\RequirePackage{fix-cm}
\documentclass{svjour3}                     
\smartqed  
\usepackage{graphicx}
\usepackage{cite}
\usepackage{array,multirow,makecell}
\usepackage{amsmath,amssymb,fancybox}
\usepackage[ruled,vlined,linesnumbered]{algorithm2e}
\usepackage{lineno,hyperref}
\usepackage[bottom]{footmisc}
\usepackage{float}
\usepackage{algorithmic}
\begin{document}

\title{Centrality in Modular Networks}


\author{Zakariya Ghalmane  \and Mohammed El Hassouni  \and Chantal Cherifi \and Hocine Cherifi}


\institute{Zakariya Ghalmane \at
              LRIT URAC No 29, Faculty of Science, Rabat IT center, Mohammed V University, Rabat, Morocco\\
              \email{zakaria.ghalmane@gmail.com}           
           \and
           Mohammed El Hassouni \at
              LRIT URAC No 29, Faculty of Science, Rabat IT center, Mohammed V University, Rabat, Morocco\\
              \email{mohamed.elhassouni@gmail.com}           
              \and
           Chantal Cherifi \at
              DISP Lab, University of Lyon 2\\
              \email{chantal.bonnercherifi@univ-lyon2.fr}           
              \and
              Hocine Cherifi \at
              LE2I UMR 6306 CNRS, University of Burgundy, Dijon, France\\
              \email{hocine.cherifi@u-bourgogne.fr}           
}

\date{Received: date / Accepted: date}

\maketitle

\begin{abstract}
Identifying influential nodes in a network is a fundamental issue due to its wide applications, such as accelerating information diffusion or halting virus spreading. Many measures based on the network topology have emerged over the years to identify influential nodes such as Betweenness, Closeness, and Eigenvalue centrality. However, although most real-world networks are modular, few measures exploit this property. Recent works have shown that it has a significant effect on the dynamics on networks. In a modular network, a node has two types of influence: a local influence (on the nodes of its community) through its intra-community links and a global influence (on the nodes in other communities) through its inter-community links. Depending of the strength of the community structure, these two components are more or less influential. Based on this idea, we propose to extend all the standard centrality measures defined for networks with no community structure to modular networks. The so-called "Modular centrality"  is a two dimensional vector. Its first component quantifies the local influence of a node in its community  while the second component quantifies its global influence on the other communities of the network. In order to illustrate the effectiveness of the  Modular centrality extensions, comparison with their scalar counterpart are performed in an epidemic process setting. Simulation results using the Susceptible-Infected-Recovered (SIR) model on synthetic networks with controlled community structure allows getting a clear idea about the relation between the strength of the community structure and the major type of influence (global/local). Furthermore, experiments on real-world networks demonstrate the merit of this approach.
\end{abstract}
\BlankLine
\BlankLine
\section{Introduction}
Identifying the most influential nodes in a network has gained much attention among researchers in the recent years due to its many applications. Indeed, they play a major role in controlling the epidemic outbreak \cite{vac}, increasing the publicity on a new product \cite{b2}, controlling the rumor spreading \cite{b3}. The most popular approach to uncover these central nodes is to quantify their influence using centrality measures. Various centrality measures have been proposed to quantify the influence of nodes based on their topological properties. Degree centrality, betweenness centrality, closeness centrality are one of the most basic centrality measures, yet they are widely used.

In real-world networks, community structure is a very common and important property \cite{c1,c2,c3,c4}. Although there is no formal definition of a community, it is often apprehended as a group of nodes tightly connected that are loosely connected with nodes in other communities. Previous works have shown that community structure has important effect on the spreading process in networks \cite{stra1,stra2,stra3,stra4}. However, classical centrality measures \cite{centralities} do not take into account the influence of this major topological property on the spreading dynamics. In a modular network, we can distinguish two types of links that support the diffusion process: the links that connect nodes belonging to the same community (intra-community links or strong ties) and the links that bridge the communities (inter-community links or weak ties). The former exercise a local influence on the diffusion process (i.e, at the community level), while the latter have a global influence (at the network level). Therefore, we believe that this two types of links should be treated differently. Indeed, the intra-community links contribute to the diffusion in localized densely connected areas of the networks, while the inter-community links allows the propagation to remote areas of the network. Suppose that an epidemic starts in a community, as it is highly connected, the intra-community links will tend to confine the epidemic inside the community, while the inter-community links will tend to propagate it to the other communities. As their role is quite different, we propose to represent the centrality of modular networks by a two dimensional vector where the first component quantifies the intra-community (or local) influence and the second component quantifies the inter-community (or global) influence of each individual node in the network. To compute these components, we need to split the original network into a local and a global network. The local network is obtained by removing all the inter-community links from the original network. The global network is obtained by removing all the intra-community links from the original network. Note that if the original network is made of a single connected component the global and local networks split into many connected components. Therefore, care must be taken to adapt the centrality definition to networks with multiple components. In the following, we restrict our attention to non-overlapping community structure (i.e a node belongs to a single community). Furthermore, we consider undirected and unweighted networks for the sake of simplicity, but results can be easily extended to more general situations. 

The proposed approach can be summarized as follows:
\begin{itemize}
\item Choose a standard centrality measure.
\item Compute the local network by removing all the inter-community links from the modular network.
\item Compute the Local component of the Modular centrality using the standard centrality.
\item Compute the global network by removing all the intra-community links from the modular network.
\item Compute the Global component of the Modular centrality using the standard centrality.
\end{itemize}
As nodes need to be ranked according to their centrality values, it is necessary to adopt a strategy based on a combination of the two components of the Modular centrality. Various strategies, that include different levels of information about the community structure, may be used. As our main concern, in this paper, is to highlight the multidimensional nature of centrality in modular networks rather than devising optimal ranking methods, elementary strategies are evaluated. Two straightforward combination strategies  (the modulus and the tangent of the argument of the Modular centrality) and a weighted linear combination of the components of the Modular centrality are investigated.

Experiments are conducted on modular synthetic networks in order to better understand the relative influence of the local and Global component of the Modular centrality in the propagation process. Extensive comparisons with the standard centrality measures show that Modular centrality measures provide more accurate rankings. Simulations on real-world networks of diverse nature have also been performed. As their community structure is unknown, a community detection algorithm has been used. Results confirm that node rankings based on the Modular centrality are more accurate than those made by the standard centrality measures which have been designed for networks with no community structure.

The rest of the paper is organized as follows. Related modular-based measures are discussed in the next section. In Section 3, a general definition of the Modular centrality is given. In this framework, we present the extensions to modular networks of the most influential centrality measures (closeness, betweenness and eigenvector centrality). The experimental setting is described in Section 4. We report and analyze the results of the experiments performed on both synthetic and real-world networks in Section 5. Finally, the main conclusions are presented in Section 6.

\section{Related works}

Ranking the nodes according to their centrality constitutes the standard deterministic approach to uncover the most influential nodes in a network. These measures rely usually on various network topological properties. However, the community structure of the network is rarely taken into consideration. Few researchers have payed attention to this property encountered in many real-world networks \cite{c4,stra1,stra2,stra3,stra4,stra5,stra6,stra7,stra8,stra9,stra10}. In this section, we give a brief overview of the main deterministic methods that motivates our proposition.\\

\noindent \textbf{a. Comm centrality.} N. Gupta et al. \cite{comm} proposed a degree-based centrality
measure for networks with non-overlapping community structure. It is based on a non-linear combination of the number of intra-community links and inter-community links. The goal is to select nodes that are both hubs in their community and bridges between the communities. This measure gives more importance to community bridges. Indeed, the number of inter-community  links is  raised to the power of two. Comparison have been performed with deterministic and random immunization strategies using the S.I.R epidemic model and both synthetic and real-world networks. Nodes are immunized sequentially from each community in the decreasing order of their centrality value in their respective community. Number of nodes to be removed from a community are kept proportional to the community size. Results show that the Comm strategy is more effective or at least works as well as Degree and Betweenness centrality while using only information at the community level. \\

\noindent \textbf{b. Number of neighboring communities centrality.} In a previous work \cite{notre}, we proposed  to rank the nodes according to the number of neighboring communities that they reach in one hop. The reason for selecting these nodes is that they are more likely to have a big influence on nodes belonging to various communities. Simulation results on different synthetic and real-world networks show that it outperforms Degree, Betweenness and Comm centrality in networks with a community structure of medium strength (i.e when the average number of intra-community links is of the same order than the number of inter-community links).\\

\noindent \textbf{c. Community Hub-Bridge centrality.} We also proposed the Community Hub-Bridge centrality measure in \cite{notre}. It is based on the combination of the number of intra-community links weighted by the size of the community and the inter-community links weighted by the number of neighboring communities. This measure tends to select preferentially nodes that can be considered as hubs inside large communities and bridges having many connections with various neighboring communities. According to experimental results, on both synthetic and real-world networks, this centrality measure is particularly suited to networks with strong community structure (i.e, when there is few inter-community links as compared to the number intra-community links). In this situation, it can identify effectively the most influential spreaders as compared to Degree, Betweenness, Comm and the Number of Neighboring Communities centrality measures.
A variation of this centrality measure called the Weighted Community Hub-Bridge centrality has also been introduced. It is weighted such that, in networks with well-defined community structure,
more importance is given to bridges (inter-community links), while in networks with weak community structure the hubs in the communities dominate. The goal is to target the bridges or the hubs according to the community structure strength. This measure has proved its efficiency as compared to the alternatives particularly in networks with weak community structure.\\

\noindent \textbf{d. K-sell with community centrality.} Luo et al. proposed a variation of the K-shell decomposition for modular networks \cite{luo}. They suggest that the intra-community and the inter-community links should be considered separately in the K-core decomposition process. Their method works as follows:\\
(i) After the removal of nodes with intra-community links, the K-shell decomposition of the remaining nodes is computed. It is associated with an index of $k^{W}_{core}$. \\
(ii) After the removal of nodes with inter-community links, the K-shell of the remaining nodes is computed. It is associated with an index of $k^{S}_{core}$. \\
(iii) A new measure is then calculated and assigned to each node based on the linear combination of both $k^{W}_{core}$ and $k^{S}_{core}$ in order to find nodes that are at the same time bridges and hubs located in the core of the network.\\
Experiments have been performed using SIR simulations on Facebook friendship networks at US Universities. Results show that this strategy is more efficient than the classical K-shell decomposition, the Degree and the Betweenness centrality measures.\\

\noindent \textbf{e. Global centrality.} In  \cite{evans}, M. Kitromilidis et al. propose to redefine the standard centrality measures in order to characterize the influence of Western artist. Based on the idea that 
influential artists have connections beyond their artistic movement, they propose to define the centrality of modular networks by considering only the inter-community links.  In other words, an influential artist must be related to multiple communities, rather than being strongly embedded in its own community.  Considering a painter collaboration network where edges between nodes represent biographical connections between artists, they compared Betweenness and Closeness centrality measures with their classical version. Results show that the correlation values between the standard and modified centrality measures are quite high. However, the modified centrality measures allows to highlight influential nodes who might have been missed as they do not necessary rank high in the standard measures.\\ 

All these works suggest that it is of prime interest to disentangle the local influence from the global influence in order to characterize a node centrality in modular networks. Indeed, these complementary types of influence may carry very different meanings and be more or less important in different situations. This is the reason why we propose to exploit the mesoscopic granularity level in order to extend the definition of the centrality measures that are agnostic about the community structure to modular networks. We propose to represent the centrality measures in modular networks as a two dimensional vector made of its local and Global component. If needed, these two components can be merged in a scalar value, but the combination can be made in multiple ways according to complementary available information about the network nature and topological properties.

\section{Modular centrality}

Our main objective is to take into account the community structure in order to identify influential nodes. Indeed, in modular networks, a node has two types of influence: a local influence which is linked to its community features and a global influence related to its interactions with the other communities.   Under this assumption, we provide a general definition of centrality in modular networks. We design a generic algorithm for computing the centrality of a node under this general definition. The Modular centrality extension can be naturally inferred from the various existing definitions of centrality designed for networks without community structure. To illustrate this process, we give the modular extensions of the most influential centrality measures (Betweenness, Closeness, and Eigenvector).

\subsection{ Definitions}

\subsubsection{Local component of the Modular centrality} 
Let's consider a network denoted as $G(V,E)$, where $V=\{v_{1},v_{2},....,v_{n}\}$ and $E=\{(v_{i},v_{j}) \setminus v_{i},v_{j} \in V\}$ denotes respectively the set of vertices and edges. Its non-overlapping community structure  $\mathcal{C}$ is a partition into a set of communities $\mathcal{C}=\{C_{1},..C_{k}.,C_{m}\}$ where $C_{k}$ is the $k^{th}$ community and \textit{m} is the number of communities. The local network $G_{l}$ is formed by the union of all the disjoint modules of the network $G_{l} = \bigcup_{k=1}^{m} C_{k}$.  These components are obtained by removing all the inter-community links between modules from the original network $G$. Each module represents a community $C_{k}$ denoted as $C_{k}(V_{k},E_{k})$. Where $V_{k}=\{ v_{i}^{k} \setminus v_{i} \in V \}$ and $E_{k}=\{(v_{i}^{k_{1}},v_{j}^{k_{2}}) \setminus v_{i},v_{j} \in V$ and $k_{1}=k_{2} \}$, while $v_{i}^{k}$ refers to any node $v_{i}$ belonging to the community $C_{k}$. 

For a selected centrality measure $\beta$, we define $\beta_{L}(v_{i}^{k})$ as the local centrality of the node $v_{i} \in V_{k}$. It is computed separately in each module $C_{k}$ of the local graph $G_{l}$ .\\

\subsubsection{Global component of the Modular centrality} 
Let's consider the network $G(V,E)$, the global network $G_{g}$ is formed by the union of all the connected components of the graph that are obtained after removing all the intra-community links from the original network $G(V,E)$. Let's suppose that $\mathcal{S}=\{S_{1},..S_{q}..,S_{p}\}$ is the set of the revealed connected components and $p=|\mathcal{C'}|$ is the size of the set $\mathcal{S}$, the global network is defined by $G_{g} = \bigcup_{q=1}^{p} S_{q}$. Each component  $S_{q}$ is denoted as $S_{q}(V_{q},E_{q})$. Where $V_{q}=\{ v_{i}^{q} \setminus v_{i} \in V \}$ and $E_{q}=\{(v_{i}^{q_{1}},v_{j}^{q_{2}}) \setminus v_{i},v_{j} \in V$ and $q_{1}=q_{2} \}$, while $v_{i}^{q}$ refers to any node $v_{i}$ belonging to the component $S_{q}$. In this network, there may be some isolated nodes (i.e, nodes that are not linked directly to another community). These nodes are removed from $G_{g}$ in order to obtain a trimmed network formed only by nodes linked to different communities by one hop. Consequently, the set of nodes of $G_{g}$ is defined then by $V_{g}=\{ v_{i} \in V \setminus |\mathcal{N}_{v_{i}}^{1}|\not= 0\}$. Where $\mathcal{N}_{v_{i}}^{n}$ is the neighborhood set of nodes reachable in $n$ hops. It is defined by $\mathcal{N}_{v_{i}}^{n}=\{v_{j} \in V \setminus v_{i} \not= v_{j}$ and $ d_{G}(v_{i},v_{j})<= n\}$, $d_{G}$ is the geodesic distance.
 
For a selected centrality measure $\beta$, we define $\beta_{G}(v_{i}^{q})$ as the global centrality of the node $v_{i} \in V_{q}$. 
It is computed over each connected component $S_{q}$ included in the global graph $G_{g}$. 
Remember that the global centrality measure of the removed isolated nodes is set to 0.\\  
\subsubsection{Modular centrality} 
It is a vector with two components. The first component quantifies the local influence of the nodes in their own community  through the local graph $G_{l}$, while the second component measures the global influence of the nodes on the other communities trough the  connected components of the global graph $G_{g}$. The Modular centrality of a node $v_{i}$ is given by:
\begin{equation}
B_{M}(v_{i})= (\beta_{L}(v_{i}^{k}), \beta_{G}(v_{i}^{q})) \qquad \text{} \;\; k \in \{1,...,m\} \;\; \text{and} \;\; q \in \{1,...,p\}
\end{equation} 

Where $\beta_{L}$ and $\beta_{G}$ represent respectively the local and global centrality of the node $v_{i}$.
\\


\subsection{Algorithm}
The Modular centrality is computed as follows:\\
Step 1. Choose a standard centrality measure $\beta$.\\ 
Step 2. Remove all the inter-community edges from the original network $G$ to obtain the set of communities $\mathcal{C}$ forming the local network $G_{l}$.\\
Step 3. Compute the local measure $\beta_{L}$ for each node in its own community.\\
Step 4. Remove all the intra-community edges from the original network to reveal the set of connected components $\mathcal{S}$ formed by the inter-community links.\\
Step 5. Form the global network $G_{g}$ based on the union of all the connected component. Isolated nodes are removed from this network and their global centrality value is set to 0.\\
Step 6. Compute the global measure $\beta_{G}$ of the nodes linking the communities based on each component of the global network.\\
Step 7. Add $\beta_{L}$ and $\beta_{G}$ to the Modular centrality vector $B_{M}$.\\
The pseudo-code of the algorithm to compute the Modular centrality is given in \autoref{al1} .

\begin{algorithm}
\DontPrintSemicolon
\SetAlgoLined
\SetKwInOut{Input}{Input}\SetKwInOut{Output}{Output}
\Input{Graph $G(V,E)$, Centrality measure $\beta$}
\Output{A map $M(node: centrality \; vector)$}

\BlankLine
Remove all the inter-community links from $G$ to form the local network $G_{l}$\\
Remove all the intra-community links from $G$ to form the global network $G_{g}$\\
Create and initialize an empty map $M(node: B_{M})$\\
$B_{M}(v_{i})=(\beta_{L}(v_{i}^{k}),\beta_{G}(v_{i}^{q}))$ represent the centrality vector, where each node $v_{i}$ of the network should be associated with its local and global value according to the selected centrality measure $\beta$\\
\BlankLine
\For{each $C_{k}  \subset G_{l}$, where $k \in \{1,..., m\}$}{
\BlankLine
\For{each $v_{i}^{k} \in V_{k}$}{
\BlankLine
Calculate $\beta_{L}(v_{i}^{k})$\\
$M.add(v_{i},B_{M}(v_{i}))$
}
}
\BlankLine
\For{each $S_{q}  \subset G_{g}$, where $q \in \{1,..., p\}$}{
\BlankLine
\For{each $v_{i}^{q} \in V_{q}$}{
\BlankLine
Calculate $\beta_{G}(v_{i})$\\
$B_{M}(v_{i}).add(\beta_{G}(v_{i}))$
}
}
\BlankLine
\For{each $v_{i} \in V$}{
\BlankLine
$M.add(v_{i},B_{M}(v_{i}))$
}
\BlankLine
Return the map $M$

\caption{Generic computation of the Modular centrality}
\label{al1}
\end{algorithm}

\begin{figure*}
\begin{center}
\includegraphics[width=16cm,height=4cm]{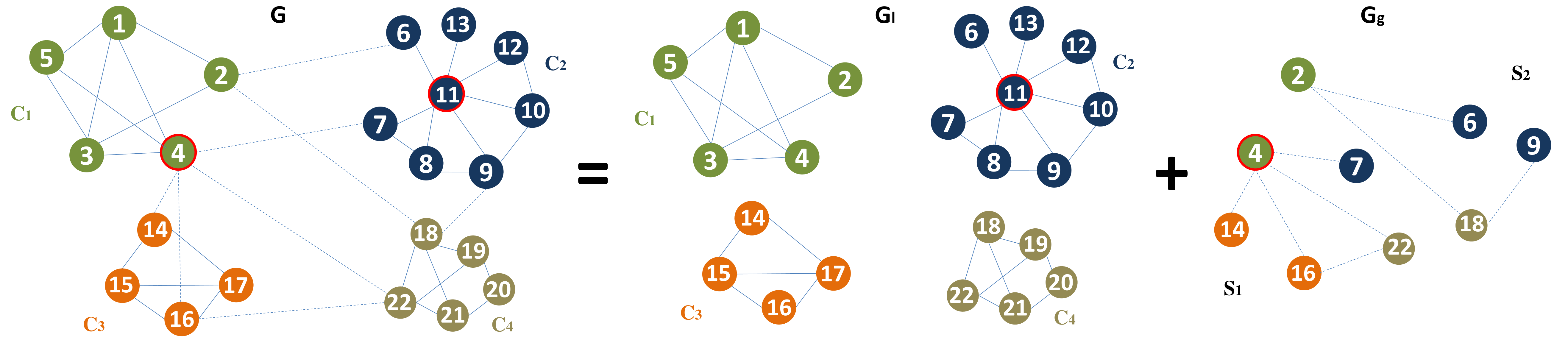}
\caption{ A toy example representing the Local network ($G_{l}$) and the Global network ($G_{g}$)associated to a a modular network ($G$) made of four non-overlapping communities.
\label{image}
}
\end{center}
\end{figure*}

\subsection{Modular extensions of standard centrality measures}
In order to illustrate the process allowing to extend a given centrality defined for a network without community structure to a modular network, we give as examples the modular definitions of the Betweenness, Closeness and Eigenvector centrality.
\subsubsection{Modular Betweeness centrality}
The modular Betweeness centrality takes into account separately paths that start and finish in the same community and those which starts and finish in different communities. For a given node $v_{i}$, it is represented by the following vector:

\begin{equation}
B_{M}(v_{i}) = (\beta_{L}(v_{i}^{k}),\beta_{G}(v_{i}^{q})) \qquad \text{} \;\; k \in \{1,...,m\} \;\; \text{and} \;\; q \in \{1,...,p\} 
\end{equation}
Where:
\begin{equation}
 \beta_{L}(v_{i}^{k}) = \displaystyle\sum\limits_{v_{s},v_{t} \in C_{k}}  \  \frac{\sigma_{st}(v_{i})}{\sigma_{st}}
\end{equation} 
\begin{equation}
 \beta_{G}(v_{i}^{q}) = \displaystyle\sum\limits_{v_{s},v_{t} \in S_{q}}  \  \frac{\sigma_{st}(v_{i})}{\sigma_{st}}
\end{equation}
$\beta_{L}$ measures the Betweenness centrality of nodes in their own community and $\beta_{G}$ measures the Betweenness centrality of nodes linking the communities. $\sigma_{st}$ is the number of shortest paths connecting nodes $v_{s}$ and $v_{t}$, while $\sigma_{st}(v_{i})$ is the number of shortest paths connecting nodes $v_{s}$ and $v_{t}$ and passing through $v_{i}$. 
\subsubsection{Modular Closeness centrality}
Modular Closeness centrality considers separately the shortest distances of nodes originating from the same or from another community than the starting node $v_{i}$. It is  defined as follows:

\begin{equation}
B_{M}(v_{i})=(\beta_{L}(v_{i}^{k}),\beta_{G}(v_{i}^{q})) \qquad \text{} \;\; k \in \{1,...,m\} \;\; \text{and} \;\; q \in \{1,...,p\} 
\end{equation}
Where: 
\begin{equation}
 \beta_{L}(v_{i}^{k}) =  \frac{1}{\sum\limits_{v_{j} \in C_{k}}  d_{ij}} 
\end{equation} 
\begin{equation}
 \beta_{G}(v_{i}^{q}) =  \frac{1}{\sum\limits_{v_{j} \in S_{q}}  d_{ij}} 
\end{equation}
$\beta_{L}$ and $\beta_{G}$ measure respectively the local and Global component of the Modular Closeness centrality. $d_{ij}$ is the length of the geodesic from node $v_{i}$ to node $v_{j}$.
\subsubsection{Modular Eigenvector centrality}
Modular Eigenvector takes into account separately both the number and the importance of the neighbors belonging to the same community and those belonging to different communities to measure its centrality. 
The community based Eigenvector  of a node $v_{i}$ is defined by the following vector:

\begin{equation}
B_{M}(v_{i})=(\beta_{L}(v_{i}^{k}),\beta_{G}(v_{i}^{q})) \qquad \text{} \;\; k \in \{1,...,m\} \;\; \text{and} \;\; q \in \{1,...,p\} 
\end{equation}
Where:
\begin{equation}
\beta_{L}(v_{i}^{k}) =  \frac{1}{\lambda}  \displaystyle\sum\limits_{v_{j} \in C_{k}} a_{ij} \beta_{L}(v_{j}^{k})
\end{equation} 
\begin{equation}
\beta_{G}(v_{i}^{q}) =  \frac{1}{\lambda}  \displaystyle\sum\limits_{v_{j} \in S_{q}} a_{ij} \beta_{L}(v_{j}^{q})
\end{equation} 
$\beta_{L}$ and $\beta_{G}$ measure respectively the local and Global component of the Modular Eigenvector centrality. $A=(a_{i,j})$ is the network adjacency matrix, i.e. $a_{i,j} = 1$ if vertex $v_{i}$ is linked to vertex $v_{j}$, 0 otherwise, and $\lambda$ is a constant.

\subsection{Toy example}
\begin{table}
   \centering
    \caption{Standard Degree centrality, Global and Local Component of the Modular Degree Centrality of the nodes in the toy example}
        \label{t0}	
      \begin{tabular}{lccccccccccc}
      \hline      

	Node ID & 4 & 11 & 18 & 22 & 1 & 2 & 3 & 16 & 5 & 7 & 8\\
	\hline
	$ \beta $ & \textbf{7} & \textbf{7} & 5 & 5 & 4 & 4 & 4 & 4 & 3 & 3 & 3 \\
	$ \beta_{L} $ & 3 & \textbf{7} & 3 & 3 & 4 & 2 & 4 & 2 & 3 & 2 & 3 \\
	$ \beta_{G} $ & \textbf{4} & 0 & 2 & 2 & 0 & 2 & 0 & 2 & 0 & 1 & 0 \\
	\hline
	Node ID & 9 & 10 & 14 & 15 & 17 & 19 & 21 & 6 & 12 & 20 & 13 \\
	\hline
	$ \beta $     & 3 & 3 & 3 & 3 & 3 & 3 & 3 & 2 & 2 & 2 & 1 \\
	$ \beta_{L} $ & 2 & 3 & 2 & 3 & 3 & 3 & 3 & 1 & 2 & 2 & 1 \\
	$ \beta_{G} $ & 1 & 0 & 1 & 0 & 0 & 0 & 0 & 1 & 0 & 0 & 0 \\
	\hline
      \end{tabular}%
\end{table}
The toy example reported in \autoref{image} allows to illustrate the two types of influence that can occur in a modular network. For the sake of simplicity, we consider the Degree centrality measure. In this case, $v_{4}$ and $v_{11}$ are the most influential nodes as they have the highest degree value $(\beta(v_ {11})=\beta(v_ {4})= 7)$. Even though they share the same degree value the influence they have on the other nodes of the network is not comparable. Indeed,  their position in the network are quite different:  $v_{11}$ is embedded in its community, while $v_{4}$ is at the border of its community. Inspecting the local and global networks give us a clear picture of their differences. As shown in the local network $G_{l}$, node $v_{11}$ is the most influential at the community level since it is linked to all the of nodes of its community $(\beta_{L}(v_{11}^{2})=7)$, while $v_{4}$ is only linked to 3 nodes of its community $(\beta_{L}(v_{4}^{1})=3)$. Actually,  both nodes $v_{1}$ and $v_{3}$ are more influential than  node $v_{4}$ in the community $C_{1}$ with their higher local Degree values. Looking at the Global network $G_{l}$, it appears clearly than node $v_{4}$ is the most influential node at the network level since it is connected to 4 nodes inside its component $(\beta_{G}(v_{4}^{1})=4)$.These nodes belong to all the other communities of the network ($C_{1}$, $C_{2}$ and $C_{3}$). Therefore, $v_{4}$ is more influential than node $v_{11}$ in the global network $G_{l}$ because of its ability to reach the different modules of the network as compared to $v_{11}$ which is influential only locally (in the community $C_{2}$).

To sum up, it can be noticed from this example that when we consider the Degree centrality, the community hubs are the most influential spreaders locally due to their ability to reach a high number of nodes in their own communities. The bridges which are linked to various communities are the most influential spreaders globally as they allow to reach a high number of communities all over the network.




\subsection{Modular centrality ranking strategies}
In order to rank the nodes according to their centrality, it is necessary to derive a scalar value from the Modular centrality vector. To do so, we can proceed in many different ways.  In order to highlight the essential features of centrality in modular networks, we choose to consider three strategies. The first two are straightforward.  Indeed, a simple way to combine the components of the Modular centrality is to use the modulus and the argument of this vector. The third strategy uses more information about the community structure in order to see if this can be beneficial.\\
The modulus  $r$  of the modular vector $B_{M}$ of a node $v_{i}$  is defined by: 
\begin{equation}
r(v_{i})=||B_{M}(v_{i})||=\sqrt{(\beta_{L}(v_{i}^{k}))^{2}+(\beta_{G}(v_{i}^{q}))^{2}} \qquad \text{} \;\; k \in \{1,...,m\} \;\; \text{and} \;\; q \in \{1,...,p\}
\end{equation}

The argument $\varphi$ of the modular vector $B_{M}$ of a node $v_{i}$ is defined as follows:
\begin{equation}
\varphi(v_{i})=arctan\left( \frac{\beta_{G}(v_{i}^{q})}{\beta_{L}(v_{i}^{k})}\right) \qquad \text{} \;\; k \in \{1,...,m\} \;\; \text{and} \;\; q \in \{1,...,p\}
\end{equation}

Rather than using the argument, we propose to use the tangent of the argument because it as a higher range.
It is defined by:
\begin{equation}
tan(\varphi(v_{i}))=\frac{\beta_{G}(v_{i}^{q})}{\beta_{L}(v_{i}^{k})} \qquad \text{} \;\; k \in \{1,...,m\} \;\; \text{and} \;\; q \in \{1,...,p\}
\end{equation}
Note that in these ranking strategies the information used about the community structure is very limited. As we expect that integrating more knowledge about the community structure in the combination strategy of the Modular centrality components may improve the efficiency of the ranking method, we also investigate the so-called "Weighted Modular measure". It is based on a linear combination of the components of the Modular centrality vector weighted by a measure of the strength of the communities.\\
The Weighted Modular measure $\alpha_{W}$ of a node $v_{i}$ is given by:
\begin{equation}
\alpha_{W}(v_{i})=(1-\mu_{C_{k}})*\beta_{L}(v_{i}^{k})+\mu_{C_{k}}*\beta_{G}(v_{i}^{q})
\end{equation}
Where $k \in \{1,...,m\}$, $q \in \{1,...,p\}$ and:
\begin{equation}
 \mu_{C_{k}} =   \frac{\sum_{v_{i} \in C_{k}} k^{inter}(v_{i}^{k})}{\sum_{v_{i}  \in C_{k}} k(v_{i}^{k})}
\end{equation}
Where $\mu_{C_{k}}$ is the fraction of inter-community links of the community $C_{k}$.\\ 
$k^{inter}(v_{i}^{k})$ is the number of inter-community links of node $v_{i}^{k}$ and $k(v_{i})$ is the degree of node $v_{i}^{k}$.\\
The Weighted Modular measure works as follows:
\begin{itemize}
\item A community $C_{k}$, where the intra-community links predominate is densely connected and therefore it has a very well-defined community structure. If an epidemic starts in such a cohesive community, it has more chance to stay confined than to propagate trough the few links that allows to reach the other communities of the network. In this case, priority must be given to local immunization. Consequently, more weight is given to the Local component of the Modular centrality $\beta_{L}$ to target the most influential nodes in the community since it is well separated from the other communities of the network.
\item A community $C_{k}$ where the inter-community links predominate has a non-cohesive community structure. It is more likely that an epidemic starting in this community diffuses to the other communities trough the many links that it shares with the other communities. Consequently, more weight is given to the Global component of the Modular centrality measure $\beta_{G}$ in order to target nodes that can propagate the epidemic more easily all over the network due to the loose community structure of $C_{k}$.
\end{itemize}

\section{Experimental setting}
In this section, we give some information about the synthetic and real-world dataset used in the empirical evaluation of the centrality measures. The SIR simulation process is recalled, together with the measure of performance used in the experiments.
\subsection{Dataset}
\subsubsection{Synthetic networks}
In order to generate artificial modular networks with controlled topological properties, the LFR benchmark is used \cite{lfr}. It allows to generate small-world networks with power-law distributed degree and community size. The input parameters of the model are the number of nodes, the desired average and maximum degree, the exponents for the degree and the community size distributions, and the mixing coefficient. The mixing coefficient parameter $\mu$ value ranges from 0 to 1. It represents the average proportion of links between a node and the ones located outside of its community. This parameter allows to control the strength of the community structure. If  its value is low, there is few links between the communities and they are well separated from each other. A high value of $\mu$ indicates a very loose community structure. Indeed, in this case, a node share more links with nodes outside its community than with nodes inside its community. Experimental studies have shown that typical value of the degree distribution exponent in real-world networks vary in the range $2 \leq \gamma \leq 3$. Networks can have different size going from tens to millions of nodes. In addition, it is also difficult to characterize the average and the maximal degree since they are very variable. Consequently, we choose for these parameters some consensual values while considering also the computational aspect of the simulations. They are reported in \autoref{t1}.\\

 \begin{table}
   \centering
    \caption{LFR network parameters}
        \label{t1}	
      \begin{tabular}{lc}
      \hline      

	Number of nodes & 4000 \\
	Average degree & 7 \\
	Maximum degree & 80 \\
	Exponent for the degree distribution & 2.8 \\
	Exponent for the community size distribution & 2 \\
	Mixing parameter & 0.1, 0.4, 0.7 \\
	Community size range & [15 200] \\
	\hline
      \end{tabular}%
\end{table}

\subsubsection{Real-world networks}
Although the LFR model produces pretty realistic networks, uncontrolled properties such as transitivity and degree correlation can deviate significantly from those observed in real-world networks \cite{lfrgunce}. Therefore, it is necessary to use real-world networks in the evaluation process. In order to cover a wide range of situations, we selected networks from various origin: online social networks, collaboration networks, technological networks, communication networks. All networks are undirected and unweighted. Experiments are performed on their largest connected component. The Louvain Algorithm is used to unveil the community structure of these networks. We choose this greedy optimization method for its simplicity. Furthermore, this popular algorithm has proved to be a good compromise between efficiency and complexity when used in many different type of networks \cite{gunce2} \cite{gunce3}.

\textbf{- Social networks:}
Four Samples of the Facebook Network are used. The ego-Facebook network collected from survey participants using the Facebook app. \footnote{\url{http://snap.stanford.edu/data}\label{first}} \cite{egoface} and the Facebook friendship network at 3 US universities (Caltech, Princeton, Georgetown) collected by Traud et al \cite{traud}. Nodes represent individuals (survey participant or members of the University), and edges represent online friendship links between two individuals. In the University network, in order to obtain data that are relevant for the spread of epidemic infections, only the relationship of individuals who live in the same dormitory or study the same major are considered. 

\textbf{- Communication network:} The Email-Eu-core\footref{first} network has been generated using email data from a large European research institution. The dataset contains only communication between institution members. Each node corresponds to an email address and an edge is established between two nodes $u$ and $v$, if at least one email has been exchanged between address $u$ and address $v$. 

\textbf{- Technological network:}  Power-Grid\footnote{\url{http://www-personal.umich.edu/~mejn/netdata/}} is a network containing information about the topology of the Western States Power Grid of the United States. An edge represents a power supply line. A node is either a generator, a transformator or a substation. 

\textbf{- Collaboration network:} GR-QC\footref{first} (General Relativity and Quantum Cosmology) collaboration network has been collected from the e-print arXiv and covers scientific collaborations between authors of papers submitted to the General Relativity and Quantum Cosmology category. If an author $i$ co-authored a paper with author $j$, the graph contains an edge from $i$ to $j$. If the paper is co-authored by $k$ authors this generates a completely connected (sub)graph on $k$ nodes. 

The basic topological properties of these networks are given in \autoref{t2}.

\begin{table}
   \centering
    \caption{Description of the structural properties of the real-world networks. \textit{N} is the total numbers of nodes, \textit{E} is the number of  edges. $<k>$, $k_{max}$ are respectively the average and the max degree. $C$ is the average clustering coefficient. $\alpha_{th}$ is the epidemic threshold of the network.}
    \label{t2}	
      \begin{tabular}{lccccccc}
      \hline

	Network & \textit{N} & \textit{E} & $<k>$ & $k_{max}$ & $C$ & $\alpha_{th}$\\
	\hline
	ego-Facebook & 4039 & 88234 & 43.69 & 1045 & 0.605 & 0.009 &\\
	Caltech & 620 & 7255 & 43.31 & 248 & 0.443 & 0.012 &\\
	Princeton & 5112 & 28684 & 88.93 & 628 & 0.298 & 0.006 &\\
	Georgetown & 7651 & 163225 & 90.42 & 1235 & 0.268 & 0.006 &\\
	Email-Eu-core & 986 & 25552 & 33.24 & 347 & 0.399 & 0.013 &\\
	Power grid & 4941 & 6594  & 2.66 & 19 &  0,107 & 0.092 &\\
	CR-QC & 4158 & 13428 & 5.53 & 81 & 0.529 & 0.059 &\\
	\hline
      \end{tabular}%
\end{table}

\subsection{SIR simulations}
To evaluate the efficiency of the centrality measures, we consider an epidemic spreading scenario using the Susceptible-Infected-Recovered (SIR) model \cite{sir}. In this setting, nodes can be classified into three classes: S(Susceptible), I(Infected) and R(Recovered). Initially, all nodes are set as susceptible nodes. Then, a given fraction $f_{0}$ of the top-ranked nodes according to the centrality measure under test are set to the state infected. After this initial setup, at each iteration, each infected node affects one of its susceptible neighbors with probability $\alpha$. Besides, the infected nodes turn into recovered nodes with the recover probability $\sigma$.
To better characterize the spreading capability, the value of the transmission rate $\alpha$ is chosen to be greater than the network epidemic threshold $\alpha_{th}$ given by \cite{epth}: 
\begin{equation}
\alpha_{th} = \frac{<k>}{<k^{2}>-<k>}
\end{equation}
Where $<k>$ and $<k^{2}>$ are respectively the first and second moments of the degree distribution. The epidemic threshold values $\alpha_{th}$ for the networks used in this study are reported in \autoref{t2}. In all the experiments, we use the same value of the transmission rate ($\alpha = 0.1$). Naturally, it is much larger than the values of the epidemic threshold $\alpha_{th}$ of all the dataset. The value of  the recover probability is also constant ($\sigma = 0.1$). We choose this small value so that each infected node may have many chances to infect its neighbors with the probability $\alpha$ before changing to the recovered status. The process continues until there is no more infected node in the network. Finally, when the spreading process stops, the number of nodes in the state "Recovered" $R$ is used to measure the spreading efficiency of the fraction of the initially infected nodes. The larger the value of $R$, the more influential the initially selected nodes. To ensure the effectiveness of the evaluation process, results of the SIR simulations are averaged over 200 independent realizations. A more detailed description of the SIR simulation process is given in \autoref{al2}.


\begin{algorithm}
\DontPrintSemicolon
\SetAlgoLined
\SetKwInOut{Input}{Input}\SetKwInOut{Output}{Output}
\Input{Graph $G(V,E)$,\\ Centrality measure: $\beta$,\\ Fraction of the initial spreaders: $f_{0}$,\\ Transmission rate: $\alpha$,\\ Recovery rate: $\sigma$,\\ The number of simulations: $n$}
\Output{The average number of the recovered nodes after the SIR simulations: $r_{av}$,\\
The standard deviation: $r_{dev}$}

\BlankLine
Rank the nodes according to the centrality measure $\beta$\\
Set all nodes as susceptible nodes\\
Compute $n_{I}$ the number of initially infected nodes: $n_{I} \leftarrow card(V) * f_{0}$\\
Select $n_{I}$ of the top ranked nodes and change their state to the infected state\\
Add the initially infected nodes to the infected list  $L\_Infected$\\

Initialize the list of the numbers of recovered nodes $n_{R}$ obtained after each simulation: $L\_nbrR \leftarrow EmptyList()$\\
\For{$counter$ from $0$ to $n$}{
$n_{R} \leftarrow 0$
\BlankLine
\While{$L\_Infected \neq Null$}{
\BlankLine
Select one infected node $v$ from the infected list $L\_Infected$\\
\For{each node $v'$ neighbor of $v$}{
\BlankLine
\If{$v'$ is susceptible}{
\BlankLine
With a probability $\alpha$ set the node $v'$ as infected\\
$L\_Infected.add(v')$
}
\BlankLine
\Else{
With a probability $\sigma$ set the node $v'$ as recovered\\
$L\_Infected.remove(v')$\\
$n_{R} \leftarrow n_{R} + 1$
}
\BlankLine
}
}
$n_{R} \leftarrow n_{R} - n_{I}$\\
$L\_nbrR.add(n_{R})$\\
}
\BlankLine

Compute the average number $r_{av}$ and the standard deviation $r_{dev}$ of the recovered nodes over the $n$ simulations based on the list $L\_nbrR$\\
Return $r_{av}$, $r_{dev}$ 
\caption{Decription of the SIR simulation process}
\label{al2}
\end{algorithm}

\subsection{Evaluation criteria}
In order to compare the ranking efficiency of a centrality measure with the one obtained by a reference centrality measure, we compute the relative difference of the outbreak size. It is given by:
\begin{equation}
    \Delta r = \frac{R_{m} - R_{s}}{R_{s}}
\end{equation}
Where $R_{m}$ is the  the final number of recovered nodes of the ranking method under test, and $R_{s}$ is the final number of recovered nodes for the reference method. Thus, a positive value of $\Delta r$ indicates a higher efficiency of the method under test as compared to the reference.

\section{Experimental Results}
Extensive experiments have been performed in order to evaluate the effectiveness of the most popular Modular centrality extensions (Degree, Betweenness, Closeness and Eigenvector centrality) as compared to their standard definition. First, the local and Global component of the various Modular centrality measures are compared to their standard counterpart. Next, the three ranking methods based on the combination of the components of the Modular centrality are also evaluated. These experiments are conducted on both synthetic and real-world networks.

\subsection{Synthetic networks}

\begin{figure*}
\begin{center}
\includegraphics[width=17cm,height=15cm]{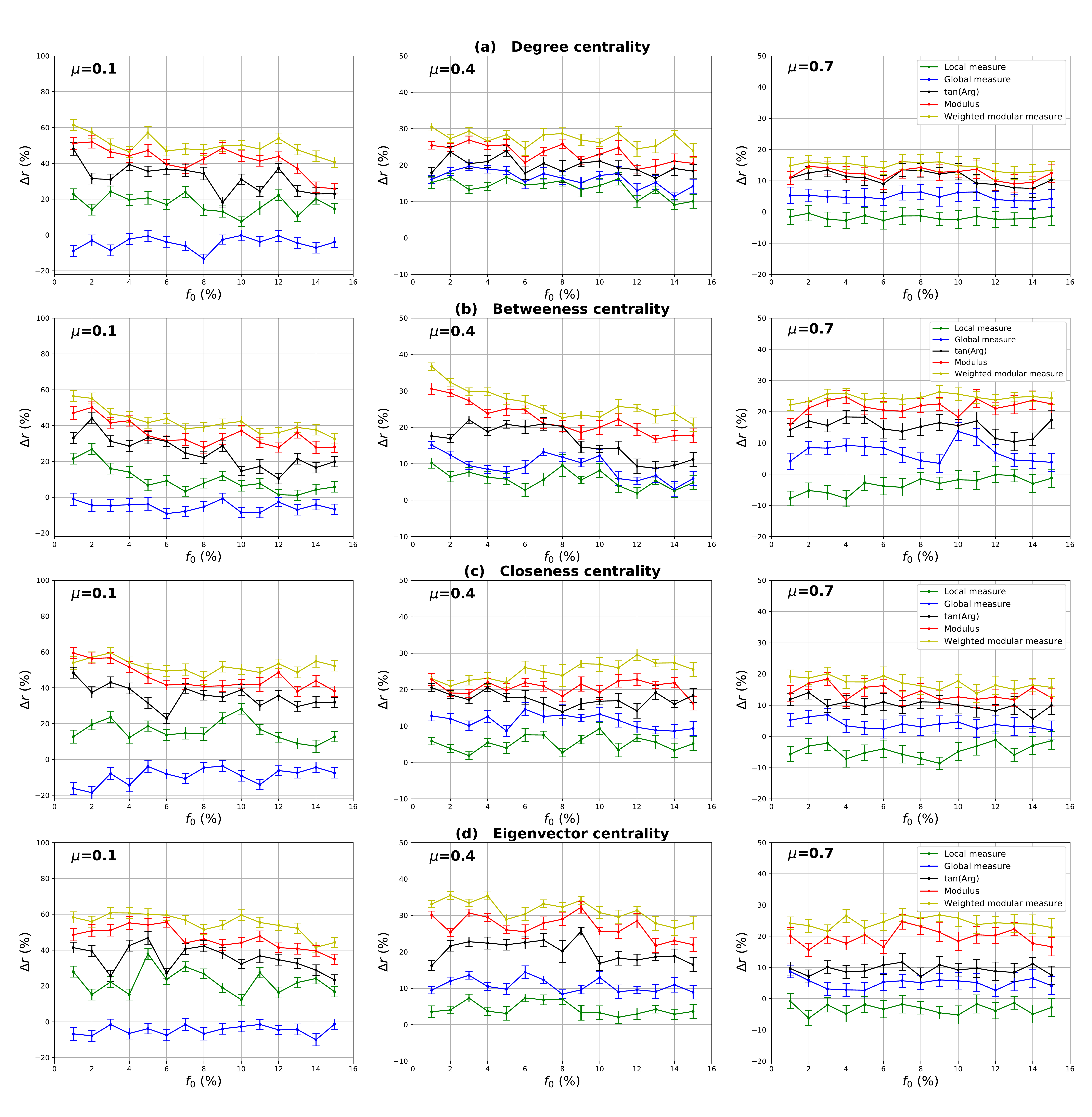}
\caption{The relative difference of the outbreak size $\Delta r$ as a function of the fraction of the initial spreaders f0 for Synthetic networks. The Degree (a), Betweenness (b), Closeness (c) and Eigenvector (d) centrality measures derived from the Modular centrality are compared to the standard counterpart designed for networks with no community structure. Networks are generated using the LFR algorithm with various $\mu$ values. For each $\mu$ value, five sample networks are
generated. The final epidemic sizes are obtained by averaging 200 SIR model simulations per network for each initial spreading coverage value. Positive $\Delta r$ value means higher efficiency of the measure under test as compared to the standard centrality. 
\label{f1}
}
\end{center}
\end{figure*}

Networks with different mixing parameter values have been generated in order to better understand the effect of the community structure strength on the  performance of the various centrality measures. \autoref{f1} represents the relative difference of the outbreak size as a function of the fraction of immunized nodes with the standard measure used as a reference. The mixing parameter values $\mu$ cover all the range of community structure strength.\\

\subsubsection{ Evaluation of the local and the Global component of the Modular centrality}
\textbf{\;\; a. Strong community structure strength}\\
 In networks with well-defined community structure, the Local component of the Modular centrality always outperforms the standard measures for all the centrality measures as it is shown in the left panels of the \autoref{f1} (when $\mu=0.1$). The gain is around 20\% as compared to the standard measure for Closeness, Degree and Eigenvector centrality. The smallest gain is for Betweenness centrality with an average value of 10\%. On the contrary, the Global component of the Modular centrality is always less performing than the standard measures. These results clearly demonstrates that it is more efficient to immunize the influential nodes inside the communities when there is few inter-community links in the networks. Indeed, as there is few inter-community edges, the infection may die out before reaching other communities. So, the local influence of nodes is more important than global influence in networks with strong community structure.\\

\textbf{b. Medium community structure strength}\\
The middle panels of \autoref{f1} show the performance of the various ranking methods in networks with community structure of medium strength ($\mu=0.4$). In this case, both the Global and Local components of the Modular centrality are always more efficient than the standard centrality. The gain in performances of the Global component of the Modular centrality is always greater than for the Local component. Indeed, the Global component outperforms the standard measure with a Gain around 12\%  for Betweenness, Closeness and Eigenvector centrality. The largest gain is for the Degree centrality with an average value of 17\%. The Local component of the Modular centrality performs better than the standard measure with a gain around 5\% for Betweenness, Closeness and Eigenvector centrality and around 12\% for Degree centrality. These results send a clear message: In networks with medium community structure strength, the global influence is more important than the local influence. Indeed, with a greater number of inter-community links, there is more options to spread the epidemics to the other communities of the network.\\

\textbf{c. Weak community structure strength}\\
The right panel of \autoref{f1} reports the comparison between the Modular centrality and the traditional centrality measures in networks with non-cohesive community structure ($\mu=0.7$). It appears that the relative difference of the outbreak size between the Global component of the Modular centrality and the standard centrality is always positive while it is always negative for the Local component of the Modular centrality. And this is true for all the centrality measures under test. In fact, there is a gain around 5\% using the Global component of the Modular centrality, while the Local component performs worse than the traditional measure with an average of 5\%  for the Degree, Betweenness, Closeness and Eigenvector centrality measures. Consequently, we can conclude that in networks with a loose community structure the global influence is dominant, even if the difference with the standard measure is not as important than for networks with a medium community structure. Indeed, in this situation ($\mu=0.7$), the inter-community edges constitute the majority of edges in the network (around 70\% of links lie between the communities). In fact, as the community structure is not well defined, minor differences are observed with a a network that has no community structure.\\   

\subsubsection{Evaluation of the ranking methods of the Modular centrality}

\autoref{f1} reports also the relative difference of the outbreak size $\Delta r$ as a function of the fraction of the initial spreaders $f_{0}$ for the three ranking methods (Modulus and Tangent of the argument of the Modular centrality, Weighted Modular measure) and for the various centrality measure and community structure strength under study. The first observation that can be made from these results is that combining the components of the Modular centrality is always more efficient than using either a single component or the conventional centrality. This remark holds for all the centrality measures studied and whatever the community structure strength. Additionally, the ranking of the three combination methods in terms of efficiency is always the same. The Weighted Modular measure ranks first. It is followed by the Modulus and then the Tangent of the argument of the Modular centrality. We believe that it is due to the fact that the Weighted Modular measure uses more information about the topology of the community structure than its alternatives. Indeed, the weights introduced in this measure allow tuning locally the relative importance of the local and Global component for each community in the network. Thus, the Weighted Modular measure can adapt to the structure of each community in the network. As a result, it is more efficient than the other proposed ranking methods. One of the main benefits of this result is to highlight the fact that significant gains can result from improving the way the local and Global component are combined and that there may be still room for improvement in this direction. In other words, even more effective measures can be obtained if relevant additional information about the community structure is used. 
Furthermore, it is noticed that the ranking strategies show their best performance in networks with well defined community structure. For instance, The Modulus of the Modular centrality outperforms the standard measure with a gain, on average, of 40\% in networks with strong community structure, 25\% in networks with community structure of medium strength and 20\% in networks with unclear community structure for all the centrality measures. For the Weighted Modular measure, the gain is around 42\% in networks with strong community structure, 29\% in networks with community structure of medium strength and 25\% in networks with unclear community structure for all the centrality measures. The gain therefore decreases as the community structure becomes more and more loose. The reason behind that behavior is that the local centrality is typical of networks with a community structure while the global centrality is also a feature  of networks with no community structure. As the mixing proportion increases, the differences with networks without community structure become less and less important. Indeed, the global network size increases until it tends to represent the major part of the original network. In the limiting case, it is a network with no community structure and the Modular centrality reduces to its Global component which is identical to the classical centrality measures.

\subsection{Real-world networks}

 \begin{table}
   \centering
    \caption{The estimated mixing parameter $\mu$ and modularity $Q$ of the real-world networks.}
        \label{t3}	
      \begin{tabular}{lcccc}
      \hline     
	Network & ego-Facebook & Power-grid & ca-GrQc & Princeton \\
	\hline
	$\mu$ & 0.03 & 0.034 & 0.095 & 0.354 \\
	$Q$ & 0.834 & 0.934 & 0.86 & 0.753\\
	\hline     
	Network & Email-Eu-core & Caltech & Georgetown\\
	\hline
	$\mu$ & 0.42  & 0.448 & 0.522\\
	$Q$ & 0.569  & 0.788 & 0.662\\
	\hline
      \end{tabular}%
\end{table}

\begin{figure*}
\begin{center}
\includegraphics[width=17cm,height=15cm]{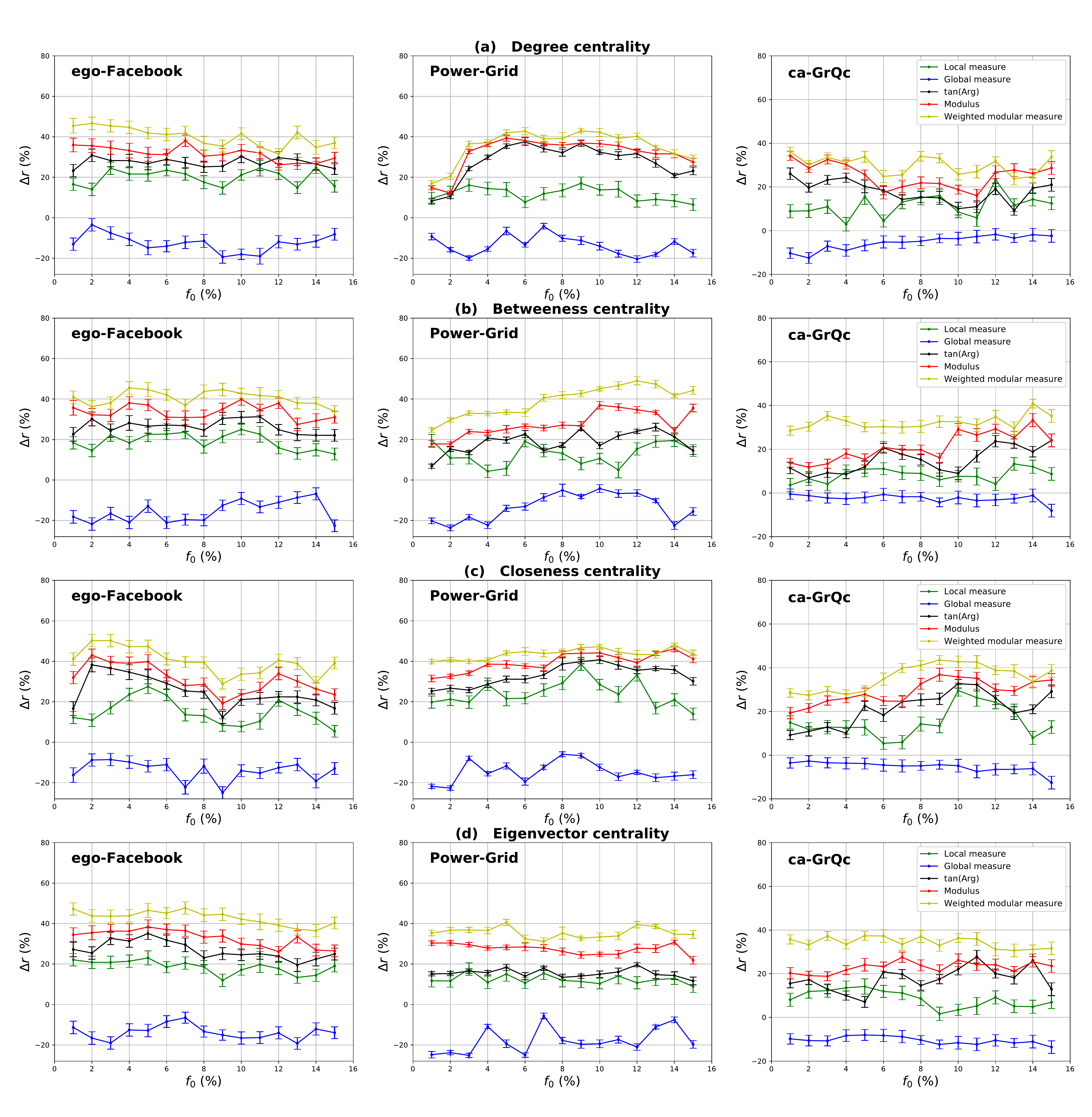}
\caption{The relative difference of the outbreak size $\Delta r$ as a function of the fraction of initial spreaders $f_{0}$. The Degree (a), Betweenness (b), Closeness (c) and Eigenvector (d) centrality measures derived from the Modular centrality are compared to the standard counterpart designed for networks with no community structure. Real-world networks with strong community structure (ego-Facebook, Power-Grid and ca-GrQc networks) are used. The estimated values of their mixing coefficient is equal respectively to 0.03, 0.034 and 0.095. 
\label{f2}
}
\end{center}
\end{figure*}

\begin{figure*}
\begin{center}
\includegraphics[width=17cm,height=15cm]{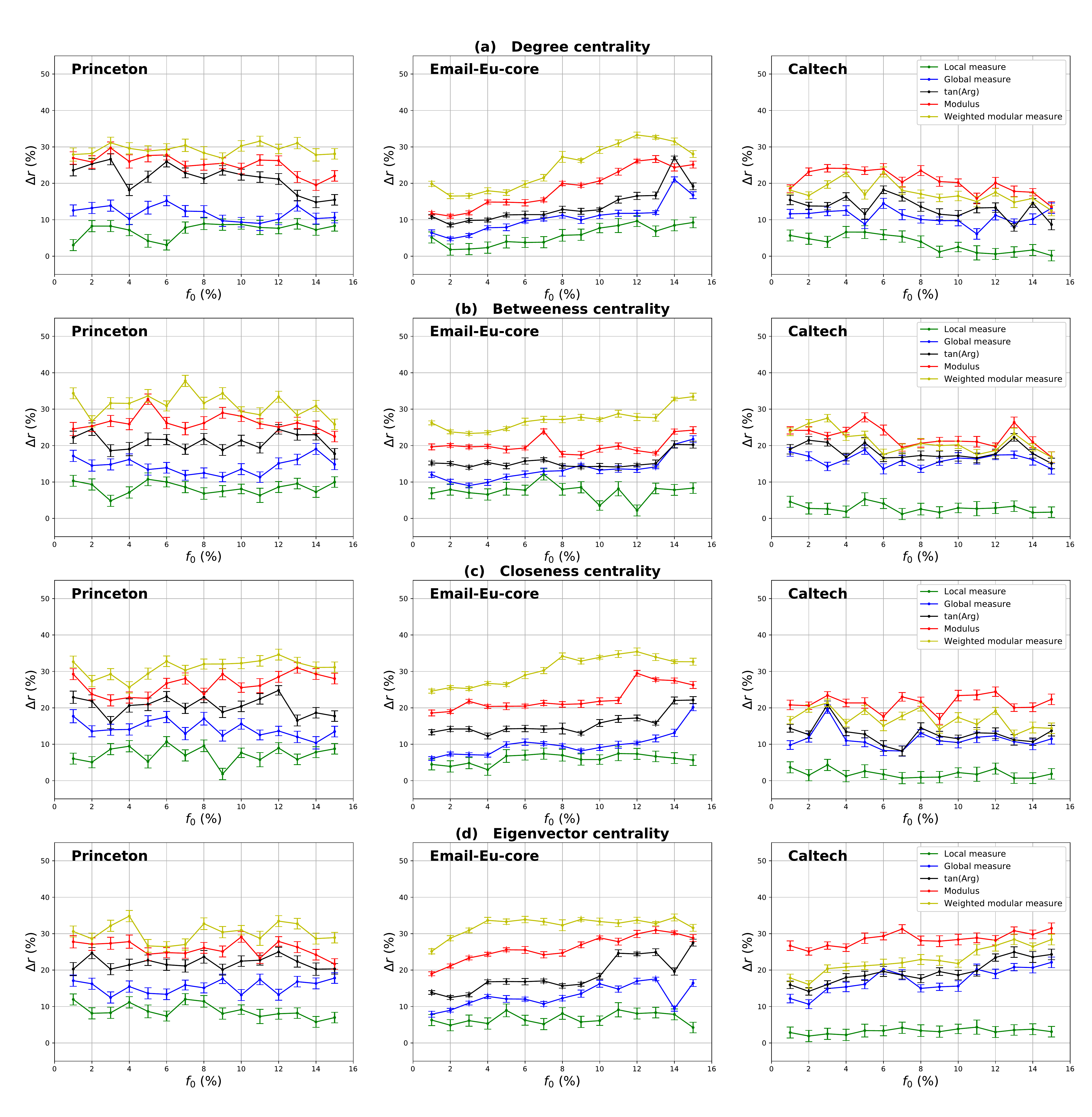}
\caption{The relative difference of the outbreak size $\Delta r$ as a function of the fraction of initial spreaders $f_{0}$. The Degree (a), Betweenness (b), Closeness (c) and Eigenvector (d) centrality measures derived from the Modular centrality are compared to the standard counterpart designed for networks with no community structure. Real-world networks with medium community structure (Princeton, Email-Eu-core and Caltech) are used. The estimated values of their mixing coefficient is equal respectively to 0.354, 0.42 and 0.44.
\label{f3}
}
\end{center}
\end{figure*}

\begin{figure*}
\begin{center}
\includegraphics[width=15cm,height=9cm]{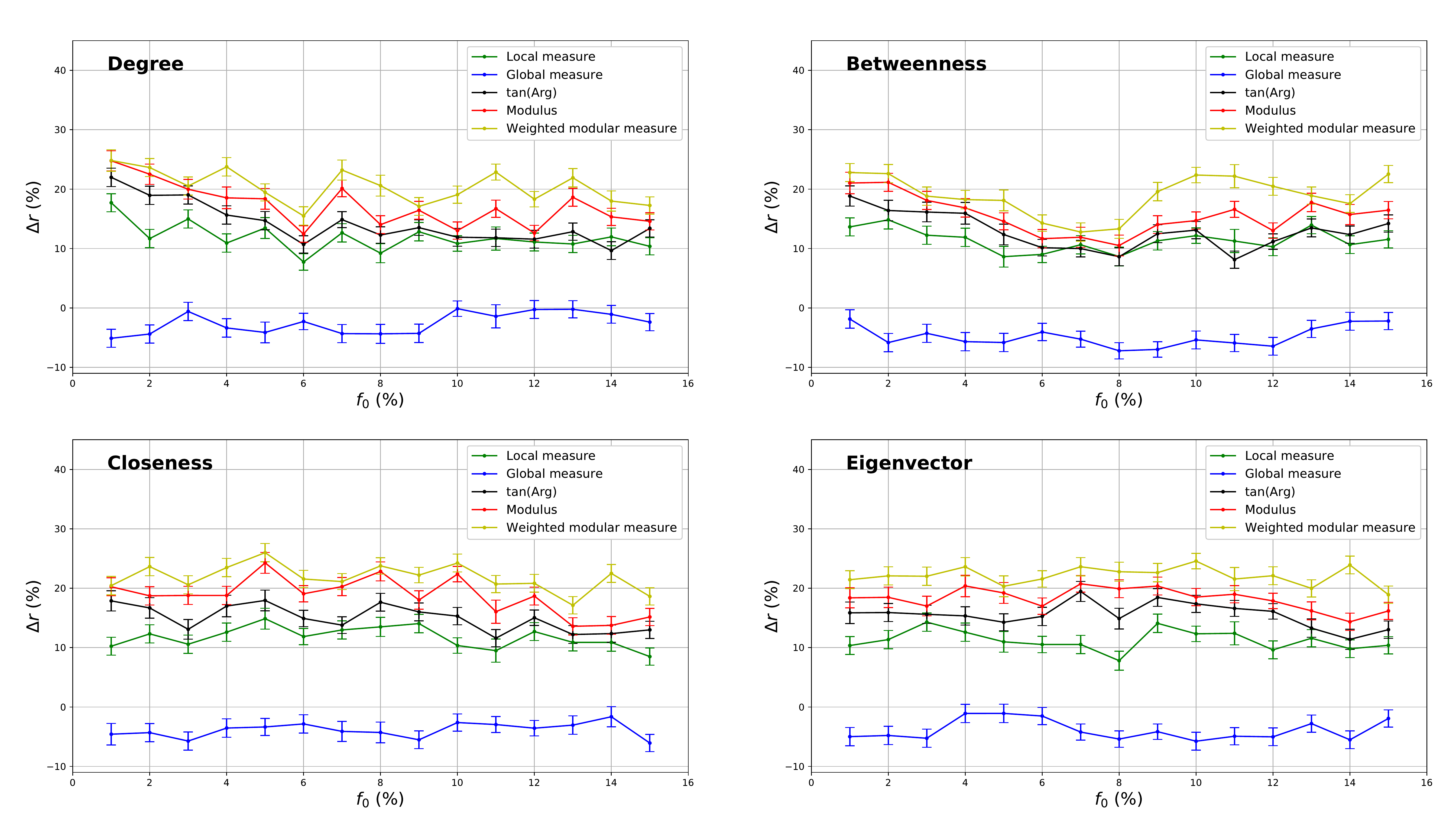}
\caption{The relative difference of the outbreak size $\Delta r$ as a function of the fraction of initial spreaders $f_{0}$. The Degree (a), Betweenness (b), Closeness (c) and Eigenvector (d) centrality measures derived from the Modular centrality are compared to the standard counterpart designed for networks with no community structure. A real-world networks with weak community structure (Georgetown) is used. The estimated value of its mixing coefficient is equal to 0.522.  
\label{f4}
}
\end{center}
\end{figure*}

In this section, we report the results of the set of experiments on real-world networks. Experiments performed with synthetic networks have shown that the community structure strength plays a major role in determining the performance of the various centrality measures. Therefore, we adopt the same presentation for real-world networks in order to link the results of this set of experiments with those obtained using synthetic networks. Once the community structure has been uncovered using the Louvain algorithm,  the mixing proportion parameter is computed for each network. Estimated values are reported in \autoref{t3}. According to these results we can classify the ego-Facebook network, Power Grid and the ca-GrQc as networks with strong community structure.  Princeton, Email-Eu-core and Caltech have a community structure of medium strength while Princeton has a weak community structure.\\

\subsubsection{Evaluation of the local and Global component of the Modular centrality}

\textbf{\;\; a. Strong community structure strength}\\
The relative difference of the outbreak size between the community based measures and the standard measure is reported in \autoref{f2}. To evaluate their performances in networks with strong community structure, ego-Facebook, power grid and the ca-GrQc networks are used. For these networks, the estimated mixing parameter value ranges from to 0.03 to 0.095. 
In this figure, we notice that for all the centrality measures under test the standard measure outperforms the Global component of the Modular centrality while it is less performing that its Local component. Let's consider for example the Betweenness centrality. With a fraction of the initial spreaders equal to 8\%, the gain in terms of the outbreak size for the Local component of the Modular centrality as compared to the standard Betweenness is  19\% for the ego-Facebook network, 14\% for  Power-Grid and and 9\% for ca-QrGc. Conversely, in the same situation, the loss associated with the use of the Global component of the Modular centrality instead of the standard Betweenness ranges from  4\% to  11\%. 

In these networks, communities are densely connected and there are few links lying between the communities. Therefore, in most cases, contagious areas are found in the core of the communities and the spread of the epidemic may stop before even reaching the community perimeter. Thus, there is a low probability that a bridge (inter-community link) propagates the epidemic to the other communities. This is the reason why the Local component of the Modular centrality performs always better than the Global component. Furthermore, we can also notice on \autoref{f2} that when the mixing parameter value increases, (i.e, the community structure gets weaker), the Local component of the Modular centrality gets less efficient while the Global component performs better. This is due to the fact that  the Global component increases with the number of inter-community links.\\

\textbf{b. Medium community structure strength}\\
\autoref{f3} shows the relative difference of the epidemic outbreak size between the community based measures and the standard centrality measure for real-world networks with medium community structure strength. For that purpose, Princeton, Email-Eu-core and Caltech networks are used since their estimated mixing parameter values range from 0.354 to 0.448. Results are very clear. In all the situations, both the local and the Global component of the Modular centrality measures outperform the standard centrality measure. In addition, there is still a slight advantage for the Global component on the Local component. To set these ideas on a simple example, let us consider the Betweenness centrality with an initial fraction of infected nodes equal to  8\%. The Global component of the modular Betweenness measure is more efficient than the traditional Betweenness with a gain of 12\% for the Princeton, 13\%  for Email-Eu-core and 15\%  for the Caltech network. These figures need to be compared to a gain of 8\% for Princeton, 7\% for Email-Eu-core and 3\% for Caltech using the Local component of the Modular Betweenness instead of the classical Betweenness centrality. One can also notice that the gap between their respective performance gets bigger as the value of the mixing parameter increases. Indeed, as the community structure gets weaker, the relative influence of the Global component of the Modular centrality becomes more and more important. In these networks, nodes have approximately as many internal links as there are external links. Therefore, the epidemic can spreads easily to all the communities in the network through the large number of inter-community links. This is the reason why the Global component of the Modular centrality outperforms always the Local component.  In addition to that, the community structure of the network is still well preserved, which explains that the Modular centrality is more efficient than the classical centrality.\\

\textbf{c. Weak community structure strength}

\autoref{f4} shows the relative difference of the epidemic outbreak size between the modular based centrality measures and the standard centrality for the Georgetown network. With a mixing parameter value equal to 0.522, this network is classified as a network with a weak community structure.  In all circumstances, the standard measure performs better than the Local component of the Modular centrality and it performs worse than its Global component. On average, there is a gain of around 10\% for the Global component compared to a loss of 5\% for the Local component of the four centrality measures under test. In this type of networks, the inter-community links predominate, which translates into a greater influence of the Global component of the Modular centrality. Indeed, the epidemic can spread more easily into the various communities of the network through the big amount of external links. Additionally, we notice that the relative difference of the outbreak size between the Global component of the Modular centrality measure and the standard measure  decreases  as compared to networks with a medium community structure. Indeed, there is less and less topological differences between networks with a weak community structure and networks that have no community structure as the value of the mixing parameter increases.\\

\subsubsection{Evaluation of the ranking methods of the Modular centrality}
 Figures 3 to 5 report also the relative difference of the epidemic outbreak size between the Modular centrality ranking methods and the standard centrality measures. The results are clear evidence of the efficiency of the Modular centrality. Whatever the ranking strategy of the Modular centrality adopted, it outperforms in all the situations the local and Global component of the Modular centrality and the classical centrality. The improvements in terms of performance compared to the classical centrality are quite significant. For instance, with a fraction of initial spreaders equal to 8\% the modulus of the Betweenness Modular centrality allows a gain of 45\% on the ego-Facebook network, 28\% on Princeton and 24\% on Georgetown. As the ranking strategies use both the local and the global information of each node, they are more efficient than measures relying on either local or global information taken separately. Furthermore, the Weighted Modular measure is usually the most efficient measure in most cases. It uses the fraction of inter-community links as an additional information to target the most influential spreaders in each community. It can give more or less weight to the local and the Global component according to the individual community structure strength. This explains its superiority over the other ranking measures.
To summarize, these experiments reveal that combining the components of the Modular centrality, allows to design efficient ranking methods. In addition, using more relevant information about the community structure at the community level allows to design even more efficient ranking methods. Moreover, the ranking measures exhibit their best results in networks with strong community structure. \\ 


\section{Conclusion}
In this paper, we propose a generic definition of centrality measures in networks with non-overlapping community structure. It is based on the fact that the intra-community and inter-community links should be considered differently. Indeed, the intra-community edges contribute to the diffusion in localized densely connected areas of the network, while the inter-community links allows the global propagation to the various communities of the network. Therefore, we propose to represent the centrality of modular networks by a two dimensional vector, where the first component measures the local influence of a node in its community and the second component quantifies its global influence on the other communities. Based on this assumption, centrality measures defined for networks with no-community structure can be easily extended to modular networks. Considering the most influential centrality measures as typical examples, we defined their modular extension. Experiments based on an epidemic spreading scenario using both synthetic and real-world networks have been conducted in order to better understand the influence of the two components of the Modular centrality. First of all, results on synthetic and real-world networks are quite consistent. It appears that the Local component is more effective in networks with a strong community structure while the Global component takes the lead as the community structure gets weaker. Comparison with the classic centrality always turns to the advantage of the Modular centrality. More precisely, in networks with strong community structure, the Local component of the Modular centrality outperforms the Global component and the standard centrality, while in networks with medium or weak community structure the Global component performs better than its alternatives. Moreover, it is also observed that combining both components of the Modular centrality in order to rank the nodes according to their influence is always more efficient than to use a single component. Furthermore, further gain can be obtained if the ranking strategy incorporates more information about the community structure strength.


\label{sec:references}

\end{document}